\documentclass[superscriptaddress,prl,twocolumn]{revtex4}
\usepackage{amsmath}
\usepackage{amssymb}
\usepackage{amscd}
\usepackage{graphicx}
\usepackage{bbm}
\usepackage{doi}
\usepackage{dsfont}
\usepackage{datetime}

\longdate

\begin{document}
\title{Defect Loops in Three-Dimensional Active Nematics as Active Multipoles}
\author{Alexander J.H. Houston}
\affiliation{Department of Physics, Gibbet Hill Road, University of Warwick, Coventry, CV4 7AL, United Kingdom.}
\author{Gareth P. Alexander}
\email{G.P.Alexander@warwick.ac.uk}
\affiliation{Department of Physics, Gibbet Hill Road, University of Warwick, Coventry, CV4 7AL, United Kingdom.}
\affiliation{Centre for Complexity Science, Zeeman Building, University of Warwick, Coventry, CV4 7AL, United Kingdom.}

\date{\today}

\begin{abstract}
We develop a description of defect loops in three-dimensional active nematics based on a multipole expansion of the far-field director and show how this leads to a self-dynamics dependent on the loop's geometric type. The dipole term leads to active stresses that generate a global self-propulsion for splay and bend loops. The quadrupole moment is non-zero only for non-planar loops and generates a net `active torque', such that defect loops are both self-motile and self-orienting. Our analysis identifies right- and left-handed twist loops as the only force and torque free geometries, suggesting a mechanism for generating an excess of twist loops. Finally, we determine the Stokesian flows created by defect loops and describe qualitatively their hydrodynamics. 
\end{abstract}
\maketitle

Active nematics are a class of materials combining self-driven, or motile, constituents with the orientational order of ordinary nematic liquid crystals~\cite{ramaswamy2010,marchetti2013,doostmohammadi2018}. Examples include bacterial suspensions~\cite{Wensink2012}, or bacteria in a liquid crystal host~\cite{zhou2014}, cell monolayers~\cite{Duclos2017} and tissues~\cite{Saw2017}, and synthetic microtubule suspensions~\cite{Sanchez2012}. The main properties are well-established for two-dimensional active nematics, including their turbulent dynamics, confinement effects and self-motile topological defects~\cite{doostmohammadi2018}. Interest is now growing in three-dimensional active nematics, with initial results on the cross-over in behaviour of defect lines as a function of cell gap~\cite{shendruk2018}, the onset of the fundamental instability in channel geometry~\cite{chandragiri2020,chandrakar2020}, the dynamics and deformations of droplets~\cite{carenza2019,ruske2021}, the characterisation and dynamics of defect loops~\cite{copar2019,duclos2020,binysh2020}, and the statistical properties of the turbulent state~\cite{krajnik2020}. Defect loops are fundamental objects in three-dimensional active nematics, analogous to $\pm 1/2$ point defects in two dimensions. They are created spontaneously and exhibit their own complex dynamics~\cite{copar2019,duclos2020}, which a local analysis sheds light on by the determination of a self-propulsion velocity for each point of the loop~\cite{binysh2020}. 

\begin{figure*}[t]
\centering
    \includegraphics[width=1\linewidth, trim = 0 10 0 5, clip]{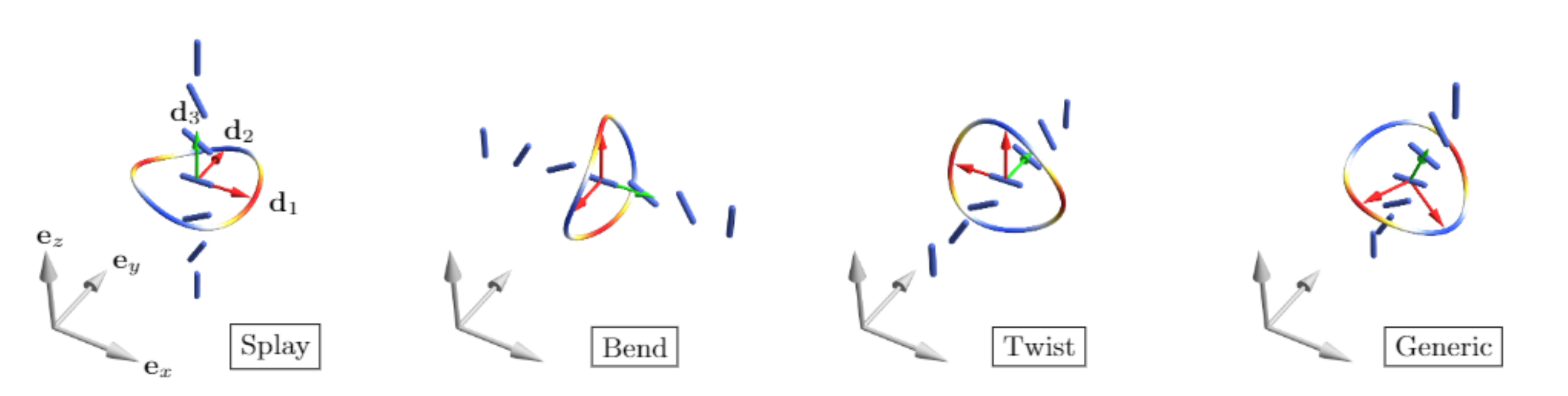}
\caption{Schematic illustration of defect loops, highlighting those of splay, bend and twist type. The defect loop is coloured according to the torsion of the curve; red where it is maximally negative and blue where it is maximally positive. The blue rods indicate the director field along a line passing through the centre of the loop and the arrows indicate the multipole frame -- in green the ${\bf d}_3$ direction of the dipole vector and in red the ${\bf d}_1$ and ${\bf d}_2$ directions coming from the quadrupole tensor. The grey basis indicates the frame defined by the plane of the director distortion ($xz$-plane), consistent across all loops and subsequent figures.} 
\label{fig:loop_structure}
\end{figure*}

The description of defects in two-dimensional active nematics as effective particles with their own dynamics has been influential in shaping understanding~\cite{giomi2013,giomi2014,khoromskaia2017,shendruk2017,cortese2018,shankar2019,angheluta2021} and it is natural to ask about the extent to which material properties in three dimensions can similarly be reduced to an effective description in terms of defect loops. Although the problem is analogous to that in two dimensions, there are differences in the topological characterisation~\cite{machon2016} and, more significantly, in the geometric diversity of defect loops. The geometry comes from both the shape of the defect loop and also the nature of the distortion in the director field around the loop. Exemplars of this come from cases where the director distortion through the middle of the loop is of pure splay, bend or twist type (see Fig.~\ref{fig:loop_structure}). These have identical properties in passive nematics with one elastic constant but behave distinctly in active systems, both in their self-propulsion dynamics~\cite{binysh2020} and in the abundances of different types, with twist defect loops found to be the most prevalent~\cite{duclos2020,binysh2020}. 

An initial analysis of active defect loops has been developed in terms of the local profile and self-propulsion velocity assigned to each point~\cite{binysh2020}. Here, we construct a complementary global description  based on an asymptotic multipole expansion for the director field. We show how the multipole structure of the active stresses generates a global self-dynamics for defect loops, involving both translational and rotational motion. The self-dynamics identifies twist loops as the only force- and torque-free states, suggesting a mechanism for the observed bias towards twist loops in three-dimensional active nematics~\cite{duclos2020}. Finally, we determine the fluid flows associated to defect loops; these are long range with a leading $1/r$ decay, such that the active hydrodynamics dominates the interactions between defect loops. We describe these qualitatively for the stable twist loops. 

A minimal model for defect loops in nematics was introduced by Friedel \& de Gennes~\cite{friedel1969}, in which the director rotates within a single plane, which here we take to define the $xz$-plane, and the director field is 
\begin{equation}
{\bf n} = \cos\theta \,{\bf e}_z + \sin\theta \,{\bf e}_x ,
\end{equation}
with $\theta$ increasing by $\pi$ as you go once around the loop. In the one-elastic-constant approximation the Frank free energy is minimised whenever $\theta$ is a harmonic function, giving $\theta = \frac{1}{4} \omega$, where $\omega$ is the solid angle function for the loop~\cite{friedel1969,Maxwell,binysh2018,binysh2020}. (In writing this we are taking the far field orientation of the director to define the $z$-direction.) This construction is independent of the shape or relative orientation of the defect loop and varying the orientation, relative to the plane of the director field, gives different geometries and local profiles for the defect loop. We illustrate this for the splay, bend and twist type loops in Fig.~\ref{fig:loop_structure}, as well as for a loop with generic orientation. 

We now define the different geometries of defect loops more precisely. The global orientation of a defect loop, and hence its geometric type, is encoded in the structure of the multipole expansion for its solid angle function. The multipole expansion is known from applications in magnetostatics~\cite{bronzan1971,gray1979} and vortex hydrodynamics~\cite{powell1964}; we state it in a form that depends only on the defect loop $K$ 
\begin{equation}
\begin{split}
\omega({\bf x}) & = \frac{1}{2} \int_{K} \epsilon_{ijk} y_j dy_k \, \partial_{i} \frac{1}{r} \\
& \quad - \frac{1}{3} \int_{K} \epsilon_{ikl} y_j y_k dy_l \, \partial_{i} \partial_{j} \frac{1}{r} + \cdots , 
%& \quad + \frac{1}{8} \int_{K} \epsilon_{ilm} y_j y_k y_l dy_m \, \partial_{i} \partial_{j} \partial_{k} \frac{1}{r} + \cdots ,
\end{split}
\label{eq:general_multipole}
\end{equation}
retaining the dipole and quadrupole terms. Here, ${\bf y}$ is a point of the loop $K$ and we are taking the position of the defect loop (${\bf x} = {\bf 0}$) to be the location of its `centre of mass' (assuming a uniform density). The dipole moment is a vector whose direction gives the principal orientation of the defect loop. The quadrupole moment is a traceless, symmetric rank 2 tensor, giving a secondary orientation. The expressions in~\eqref{eq:general_multipole} hold for any shape of defect loop, but for our current focus we calculate them explicitly only for the representative curve ${\bf y}(u) = (a \cos u , a \sin u , \frac{1}{6} \tau_0 a^2 \sin 2u)$ corresponding to a loop that is approximately a circle of radius $a$. The parameter $\tau_0$ captures the non-planarity of the loop and is the amplitude of its torsion. This yields  
\begin{equation}
\omega({\bf x}) = \pi a^2 \nabla_{{\bf d}_3} \frac{1}{r} + \frac{\pi a^4 \tau_0}{6} \,\nabla_{{\bf d}_1} \nabla_{{\bf d}_2} \frac{1}{r} + \cdots ,
\label{eq:loop_solid_angle}
\end{equation}
where $\{{\bf d}_1, {\bf d}_2, {\bf d}_3\}$ are the Cartesian basis vectors for the coordinate system adapted to the defect loop. The direction of the dipole is the basis vector ${\bf d}_3$ and its magnitude is the `area bound by the loop'. The quadrupole tensor is $\frac{\pi a^4 \tau_0}{12} [{\bf d}_1{\bf d}_2 + {\bf d}_2{\bf d}_1]$ and is proportional to $\tau_0$ so that it vanishes for planar loops without torsion. The directions ${\bf d}_1, {\bf d}_2$ correspond to those of `principal torsion'; specifically, the torsion is $\tau \approx - \tau_0 \cos 2u$ and takes its maximal negative value along the directions $\pm {\bf d}_1$ and its maximal positive value along $\pm {\bf d}_2$. This multipole frame is illustrated in Fig.~\ref{fig:loop_structure} and used in all figures. 
 
In an active nematic, the activity imparts additional material stresses $-\zeta {\bf nn}$, where $\zeta$ is a phenomenological coefficient that is positive in extensile materials~\cite{ramaswamy2010,marchetti2013,doostmohammadi2018}. For simplicity of presentation we will assume $\zeta > 0$ in what follows, as is the case in the experimental system~\cite{duclos2020}. The active stresses generate self-dynamics for the defect loop, which we characterise first by the contributions that they make to the force and torque on a spherical volume centred on the loop. In the asymptotic region ($r \gg a$) the active stress can be approximated by 
\begin{equation}
- \zeta {\bf nn} = - \zeta {\bf e}_z {\bf e}_z - \frac{\zeta \omega}{4} \bigl[ {\bf e}_z {\bf e}_x + {\bf e}_x {\bf e}_z \bigr] + \cdots ,
\end{equation} 
and using this the contribution to the force is 
\begin{equation}
\begin{split}
{\bf F} & = \int - \zeta {\bf nn} \cdot d{\bf A} , \\
& = \frac{\zeta \pi^2 a^2}{3} \Bigl[ \bigl({\bf e}_z \cdot {\bf d}_3\bigr) \,{\bf e}_x + \bigl({\bf e}_x \cdot {\bf d}_3\bigr) \,{\bf e}_z \Bigr] ,
\end{split}
\label{eq:active_force}
\end{equation}
with the integral taken over a spherical surface entirely enclosing, and centred on, the defect loop. This force depends on the surface over which the integral is taken, since the active stress is neither compactly supported nor divergence-free, but for the multipole analysis a spherical surface is natural and the result~\eqref{eq:active_force} is then independent of the radius and determined by the dipole part of the solid angle~\eqref{eq:loop_solid_angle}. Similarly, the active stress contribution to the total torque acting on the defect loop is given by 
\begin{equation}
\begin{split}
{\bf T} & =  \int {\bf x} \times \bigl( - \zeta {\bf nn} \bigr) \cdot d{\bf A} , \\
& = \frac{-\zeta \pi^2 a^4 \tau_0}{30} \biggl\{ {\bf e}_x \Bigl[ {\bf d}_1 \cdot \bigl( {\bf e}_x {\bf e}_y + {\bf e}_y {\bf e}_x \bigr) \cdot {\bf d}_2 \Bigr] \\
& \qquad + 2 {\bf e}_y \Bigl[ {\bf d}_1 \cdot \bigl( {\bf e}_z {\bf e}_z - {\bf e}_x {\bf e}_x \bigr) \cdot {\bf d}_2 \Bigr] \\
& \qquad - {\bf e}_z \Bigl[ {\bf d}_1 \cdot \bigl( {\bf e}_y {\bf e}_z + {\bf e}_z {\bf e}_y \bigr) \cdot {\bf d}_2 \Bigr] \biggr\} ,
\end{split} 
\label{eq:active_torque}
\end{equation}
where again the integral is taken over a spherical surface enclosing the entire loop and is independent of its radius; it is determined by the quadrupole part of the solid angle~\eqref{eq:loop_solid_angle}. We reiterate that the torque is proportional to the magnitude of the torsion of the loop and hence dependent on its non-planar shape. 

If the defect loop was a rigid body its self-dynamics would follow from the resistance matrix relating the force and torque to the translational and rotational velocities, however, we can also expect an internal dynamics affecting its shape and form, which in turn control the dipole and quadrupole moments. Nonetheless, even without knowledge of the internal dynamics, qualitative features of the global dynamics can be extracted from the structure of the force and torque and, in particular, the defect loop geometries for which they vanish. We consider the force first, for the three representative loops of splay, bend and twist type.

\begin{figure}[t]
\centering
    \includegraphics[width=1\linewidth, trim = 0 10 0 0, clip]{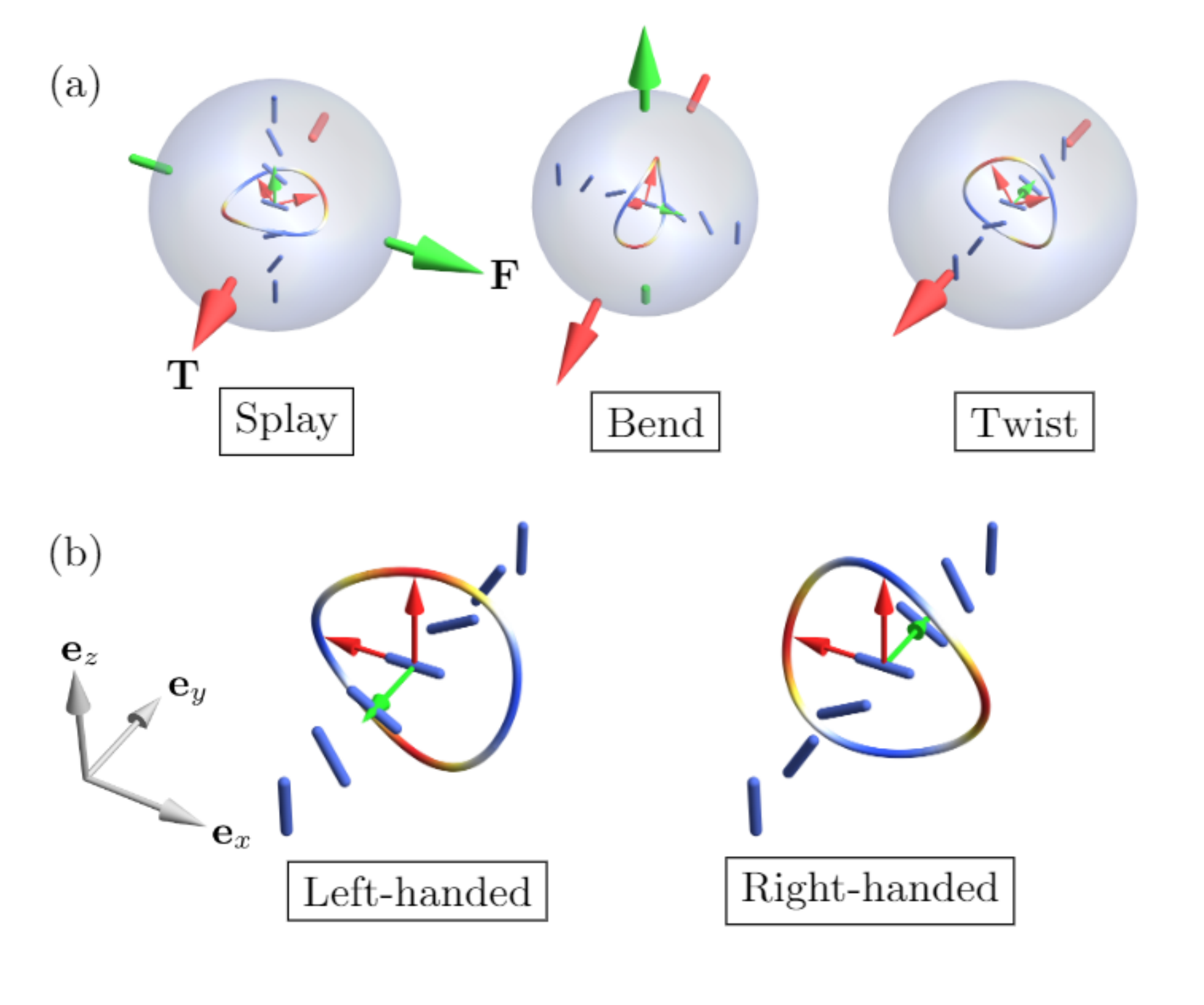}
\caption{(a) Active forces and torques experienced by generic splay, bend and twist defect loops. The external green and red arrows indicate the force ${\bf F}$~\eqref{eq:active_force} and torque ${\bf T}$~\eqref{eq:active_torque}, respectively. The torque depends on the structure of the quadrupole tensor and is shown here for representative cases. (b) Stable left- and right-handed twist loops. Note that the principal torsion directions are oppositely oriented in the two cases.} 
\label{fig:force_torque}
\end{figure}

For the splay loop (${\bf d}_3 = {\bf e}_z$) the force is directed along ${\bf e}_x$; for the bend loop (${\bf d}_3 = {\bf e}_x$) it is directed along ${\bf e}_z$; and for the twist loop (${\bf d}_3 = {\bf e}_y$) it vanishes. This is illustrated by the large green arrows in Fig.~\ref{fig:force_torque}(a). Assuming a leading diagonal response, we expect the splay loop to move along $x$, the bend loop to move along $z$ and the twist loop to remain stationary. On dimensional grounds the magnitude of the resistance should scale as $\mu a$, where $\mu$ is the visocity, so that the defect loop velocity scales as $\zeta a/\mu$ and therefore increases linearly with the size of the defect loop. In~\cite{binysh2020} the dynamics of these three geometries of defect loop were determined by assigning a local self-propulsion velocity to each point of the loop on the basis of its local director profile and exactly the same predictions obtained and confirmed by numerical solution of the full hydrodynamic equations.  

We turn now to the torque~\eqref{eq:active_torque} and the rotational motion of the defect loop. The torque is illustrated for the splay, bend and twist loops by the large red arrows in Fig.~\ref{fig:force_torque}(a). For the splay loop (${\bf d}_3 = {\bf e}_z$), taking the general orientation ${\bf d}_1 = \cos\gamma \,{\bf e}_x + \sin\gamma \,{\bf e}_y$ for the quadrupole moment (out-of-plane buckling) the torque is 
\begin{equation}
{\bf T} = - \frac{\zeta \pi^2 a^4 \tau_0}{30} \bigl[ {\bf e}_x \cos 2\gamma + {\bf e}_y \sin 2\gamma \bigr] ,
\end{equation}
and acts to reorient the defect loop from splay type to either bend or twist type. If we take the frictional resistance to scale as $\mu a^3$ on dimensional grounds, then the rotational velocity will scale as $\zeta a \tau_0/\mu$ and again increases linearly with the size of the defect loop. The situation is entirely analogous for bend loops (${\bf d}_3 = {\bf e}_x$), which experience a torque reorienting them into twist or splay geometry. For twist loops (${\bf d}_3 = {\bf e}_y$), taking a general orientation ${\bf d}_1 = \cos\gamma \,{\bf e}_z + \sin\gamma \,{\bf e}_x$ the torque is 
\begin{equation}
{\bf T} = \frac{\zeta \pi^2 a^4 \tau_0}{15} \, \sin 2\gamma \, {\bf e}_y ,
\end{equation}
and is purely about ${\bf e}_y$ so that they retain their twist character. The torque vanishes when $\gamma = 0, \frac{\pi}{2}$ and is restortative around $\gamma = \frac{\pi}{2}$, identifying this as a stable orientation for the defect loop. Of course, there are also twist loops with ${\bf d}_3 = - {\bf e}_y$; the torque they experience has a parallel description except that now the stable orientation corresponds to $\gamma = 0$ (${\bf d}_1$ parallel to ${\bf e}_z$). These two stable states differ in the handedness of the twist rotation in the director passing through the defect loop; the case ${\bf d}_3={\bf e}_y$ corresponds to right-handed twist (dextro twist loop), while ${\bf d}_3=-{\bf e}_y$ corresponds to left-handed twist (laevo twist loop). They are illustrated in Fig.~\ref{fig:force_torque}(b). 

\begin{figure*}[t]
\centering
    \includegraphics[width=1\linewidth, trim = 0 5 0 10, clip]{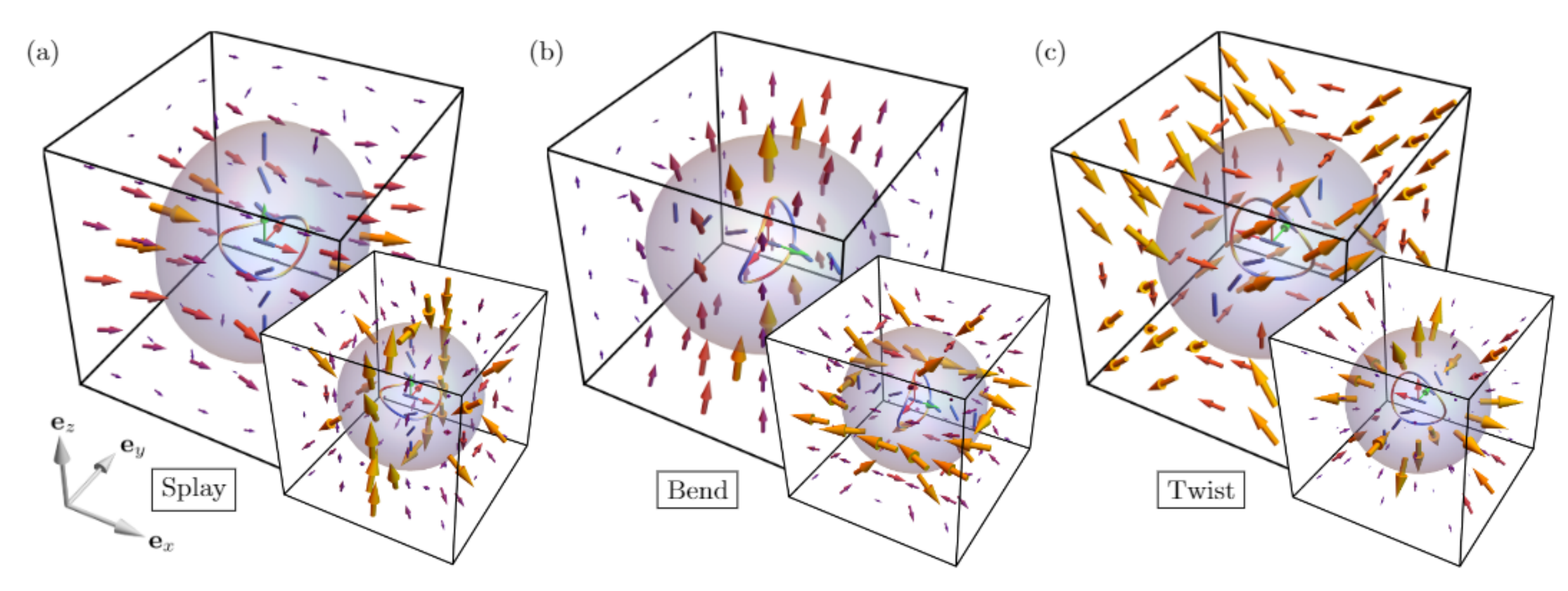}
\caption{The asymptotic flows induced by defect loops. The dipole contribution for each defect loop geometry is shown in the main panel, with the quadrupole flows shown as insets. For splay (a) and bend (b) loops the dipole and quadrupole flows show directed and rotational flows respectively, corresponding to the net active forces and torques experienced by the defect loops. (c) A stably-oriented right-handed twist loop. The stability of this configuration is reflected by the lack of propulsive or rotational flows; the dipole flow is predominantly extensional as described in the text.}
\label{fig:flows}
\end{figure*}

The existence of stable states and a general drive to convert other geometries towards these suggests that the self-dynamics will create a bias in the occurrence of different types of loops, favouring the stable twist forms. Observations in experiments and simulations~\cite{duclos2020} have found a prevalence of twist loops and it is natural to speculate that the active dynamics we have described may contribute to explaining this. In the absence of chirality (as in our analysis) one expects equal numbers of right- and left-handed twist loops, although statistics for this from experiment or simulation are not currently available. However, as the biopolymers that go into active nematics are chiral it is possible there will be an imbalance in the proportion the two types. 

We now determine the far-field structure of the fluid flows generated by defect loops and show that they confirm the self-dynamics described above. To do so, we adopt the strategy of seeking a solution of the Stokes equations with active nematic force term given by the director field of a defect loop~\cite{binysh2020}. These are $\nabla\cdot{\bf u}=0$ and 
\begin{equation}
- \nabla p + \mu \nabla^2 {\bf u} = \zeta \nabla \cdot ({\bf nn}) = \frac{\zeta}{4} \bigl[ {\bf e}_x \,\partial_{z} + {\bf e}_z \,\partial_x \bigr] \omega ,
\end{equation} 
taking the linearised form of the active stresses. The multipole expansion for the solid angle~\eqref{eq:loop_solid_angle} is given in terms of derivatives of the generating monopole $1/r$ and the resulting flow is therefore the same derivatives of the fundamental flow in response to this monopole, which provides a convenient representation for the solution~\cite{houston2021}
\begin{align}
p & = \frac{\zeta}{4} \biggl( \pi a^2 \nabla_{{\bf d}_3} + \frac{\pi a^4 \tau_0}{6} \,\nabla_{{\bf d}_1} \nabla_{{\bf d}_2} + \cdots \biggr) \frac{xz}{r^3} , \\
\begin{split}
{\bf u} & = \frac{\zeta}{16\mu} \biggl( \pi a^2 \nabla_{{\bf d}_3} + \frac{\pi a^4 \tau_0}{6} \,\nabla_{{\bf d}_1} \nabla_{{\bf d}_2} + \cdots \biggr) \\
& \quad \times \biggl\{ {\bf e}_x \biggl[ \frac{z}{r} + \frac{x^2 z}{r^3} \biggr] + {\bf e}_y \,\frac{xyz}{r^3} + {\bf e}_z \biggl[ \frac{x}{r} + \frac{xz^2}{r^3} \biggr] \biggr\} .
\end{split} \label{eq:loop_flow}
\end{align}
We remark that the fundamental solution -- the part in curly braces in~\eqref{eq:loop_flow} -- does not decay with distance; this is the active flow that would result from a localised (monopole) reorientation of the director field in an otherwise uniformly aligned nematic and the non-decay may be viewed as a signature of the fundamental instability of active nematics~\cite{simha2002}. The flow generated by a defect loop does decay but only slowly, with the leading dipole contribution falling off as $1/r$. These flows are shown in Fig.~\ref{fig:flows} for splay, bend and twist type loops. For the splay and bend loops there is a clear directed flow, consistent both with the non-zero force~\eqref{eq:active_force} and with the global self-propulsion of these loops found previously~\cite{binysh2020}. The insets in Fig.~\ref{fig:flows} show the quadrupole contribution to the flow; again, for the splay and bend loops there is a rotational character consistent with the non-zero torque~\eqref{eq:active_torque} and indicating a corresponding rotation of the defect loop. The flow generated by the stable twist loop, Fig.~\ref{fig:flows}(c), is predominantly extensional in the $xz$-plane of the director and shows neither a directed nor a rotational component. In each Cartesian plane the flow is normal to the plane and with an alternating sign in each quadrant. In the $xz$-plane this amounts to the buckling flow found in~\cite{binysh2020}. Along the $x$- and $z$-axes the flow shows local circulations, while along the $y$-axis it has a hyperbolic structure. 

The multipole flow captures the global self-propulsion and rotation, however, it does not reproduce the detailed variation in local self-propulsion velocity at each point of the loop associated with the varying director profile~\cite{binysh2020}. This suggests that a matched asymptotics between the multipole and local calculations may yield a more complete analysis.

The long-range ($\sim 1/r$) nature of the active flows generated by defect loops suggests that hydrodynamic interactions will be particularly strong and important. The strength can be compared with the elastic dipole-dipole interactions mediated by the director field~\cite{lubensky1998}, which fall off much more rapidly as $1/r^3$; indeed the quadrupole part of the velocity~\eqref{eq:loop_flow} also decays more slowly than this (as $1/r^2$) and may be expected to be more significant than the elasitic interactions. The leading character of the hydrodynamic interactions is that each loop is advected and rotated by the flow(s) generated by the other(s). As an example, we consider qualitatively the advective interactions between stable twist loops (${\bf d}_3 = \pm {\bf e}_y$). The dipole part of the flow~\eqref{eq:loop_flow} is even under ${\bf x} \to - {\bf x}$ and has different signs for the right- and left-handed loops. As a result, two twist loops of the same handedness advect each other with the same velocity, creating a collective motility reminiscent of that for pairs of scallops or dumbbells~\cite{alexander2008,lauga2008}. In contrast, two loops with opposite handedness advect each other with equal but opposite velocities. The integral curves of the dipole part of~\eqref{eq:loop_flow} form closed loops that do not visit the origin, suggesting that this contribution to the hydrodynamic interaction may lead to a periodic motion of two loops of opposite handedness -- a type of `waltzing' -- rather than simple attraction or repulsion, although it is likely that the actual dynamics will be far less regular than this heuristic picture.  

We have provided a global description of nematic defect loops in terms of a multipole decomposition of the director field. This complements the previous local analysis~\cite{binysh2020} and also shows the importance of non-planarity of defect loops in leading to net active torques. We find that twist loops are the only geometry with vanishing force and torque, providing a possible explanation of their preponderance in active nematics~\cite{duclos2020}. There are many immediate directions for development, including extending the analysis to more general shapes of defect loops, or multiple loops, and connecting this global description with the local analysis of~\cite{binysh2020} to incorporate the change in internal structure, or shape, of the defect loop to potentially develop a form of matched asymptotics. Also of interest will be to consider defect loops with non-zero topological charge and in confinement~\cite{copar2019}, and the effects of chiral active stresses~\cite{kole2021}.

%acknowledgements
This work was supported by the UK EPSRC through Grant No.~EP/N509796/1.

\end{document}